# Time Warp Edit Distance with Stiffness Adjustment for Time Series Matching

(Version 4: February 2007)

P.F. MARTEAU

**Abstract--** In a way similar to the *string-to-string correction problem* we address time series similarity in light of a *time-series-to-time-series-correction problem* for which the similarity between two time series is measured as the minimum cost sequence of "*edit operations*" needed to transform one time series into another. To define the "*edit operations*" we use the paradigm of a graphical editing process and end up with a dynamic programming algorithm that we call Time Warp Edit Distance (TWED). TWED is slightly different in form from Dynamic Time Warping, Longest Common Subsequence or Edit Distance with Real Penalty algorithms. In particular, it highlights a parameter which controls a kind of stiffness of the elastic measure along the time axis. We show that the similarity provided by TWED is a potentially useful metric in time series retrieval applications since it could benefit from the triangular inequality property to speed up the retrieval process while tuning the parameters of the elastic measure. In that context, a lower bound is derived to link the matching of time series into down-sampled representation spaces to the matching into the original space. Empiric quality of the TWED distance is evaluated on a simple classification task. Compared to Edit Distance, Dynamic Time Warping, Longest Common Subsequence and Edit Distance with Real Penalty, TWED has proved to be quite effective on the considered experimental task.

## I. INTRODUCTION

More and more computer applications are faced with the problem of searching large datasets for time series which are close to a given query element under some similarity criteria. Financial and stock data analysis [40], moving objects identification [4], astronomy [29], medicine [20], meteorology, data mining [1], time-stamped event data processing [39], network monitoring [30] are but a few of the numerous examples that could be cited.

All these applications embed time series in a representation space and exploit some similarity measure defined for this space. Similarity measures fall basically into three categories:

- Non elastic metrics such as Lp-norms that do not support time shifting, such as Euclidian

Manuscript received September 19, 2006. This work was supported in part by the French Ministère de l'enseignement supérieur et de la recherche, Plan Pluri-formation "Modélisation" Valoria, Université de Bretagne Sud.
    [1]P.F. Marteau is with the VALORIA Computer Science Lab., University of Bretagne Sud, BP573, 56017 Vannes, France, (telephone: +33 (0)2 97 01 72 99, e-mail: pierre-francois.marteau@ univ-ubs.fr).

Distance (ED) and Correlation,

- Elastic similarity measures that tolerate time shifting, but are not metrics such as Dynamic Time Warping (DTW) [36], [34] or Longest Common Subsequence (LCSS) [6], [37].

- Elastic metrics that tolerate time shifting, such as Edit distance with Real Penalty (ERP) [5].

When considering time series information retrieval, working in a metric space can be appealing, because a lot of data structures (essentially tree-based structures) and algorithms (partitioning, pivoting, etc.) have been optimized and made available for efficiently indexing and retrieving objects in metric spaces: see [3] for a review. All these structures and algorithms take advantage of the triangle inequality which allows for the efficient pruning of a large number of time series which are too far away from the query. For some non-metric measures, all these data structures can still be used if a lower bounding approximation, which needs to be a metric, is available. A lower bound of the sort exists for both LCSS and DTW as detailed in [37].

Further more, the need for processing time-stamped data (event data or data that are not sampled coherently) is becoming particularly significant [39][30] in stock analysis, network monitoring, fault analysis, etc.

In this paper we address the case of elastic metrics, namely elastic similarity measures that jointly exploit time shifting (measured using timestamps or sample indices) and possess all the properties of a distance, in particular the triangle inequality. Our contribution is basically four folded:

- The first contribution of this paper is the proposal of new elastic metric which we call TWED ("Time Warp Edit Distances"). This contribution has to be placed in the perspective of former works that seek to combine *Lp-norms* with the edit distance, in particular in the light of the ERP distance [5] that can support local time shifting while being a metric. Other elastic similarity measures that belong to the Dynamic Time Warping category are not metrics since they do not satisfy the triangle inequality. Part II of the paper promotes the need for triangle inequality to process time series in a data compression context based on a down sampling perspective.

- The second contribution is related to the introduction of a parameter we call *stiffness* which controls the *elasticity* of TWED, placing this kind of distance in between the Euclidian distance (somehow a distance with '*infinite stiffness*') and DTW (somehow a similarity measure with no '*stiffness*' at all). One of the differences between TWED and former similarity measures is the use of time stamp differences between compared samples as part of the local matching costs. The motivation for such a characteristic is also given in part II of the paper.

- The third contribution proposes a lower bound for the TWED measure which allows one to link the evaluation of the matching of two time series into down-sampled representation spaces to the evaluation of their matching into their original representation spaces.

- The fourth contribution of the paper is an empiric evaluation of the quality of TWED based on a simple classification experiment that provides some highlights on the effectiveness of TWED compared to the Euclidian Distance (ED), DTW, LCSS and ERP. The influence of the *stiffness* parameter on classification error rates is also analyzed.

The paper is organized as follows. Section II addresses the motivation aspects. Section III briefly presents the main relevant founding works on elastic distances for time series matching.

Section IV details the definition and implementation of the Time Warp Edit Distance with stiffness adjustment that is proposed in this paper. Section V details a lower bounding procedure we suggest to speed up range queries processing. Section VI describes a classification experiments that shows the empirical effectiveness of TWED comparatively to the Euclidian distance and other classical elastic measures. Section VII concludes the paper and proposes some perspectives.

II. MOTIVATION FOR A SIMILARITY MEASURE THAT VERIFIES THE TRIANGLE INEQUALITY AND TAKES TIMESTAMP DIFFERENCES INTO ACCOUNT

The use of elasticity theory to model the behaviour of non-rigid curves, surfaces, and solids as function of time has given rise to a lot of applications in medical image analysis, vision or computer graphics (see [35] [27] for surveys). These models are fundamentally dynamic and unify the description of shape and the description of motion. In another hand, elastic distances have been proposed to define similarity measures that are tolerant to object deformations, in particular stretching or shrinking. Although the analogy with physical models of deformable objects makes sense, we do not extend it too far since, in the physical sense, the laws that should govern the matching of deformable object are not always available or costly to cope with.

Our motivation for using time stamps (or sample indices) is related to the way we want to control the elasticity of the measure. Differences of indices between match samples have been successfully used to improve elastic measures such as Dynamic Time Warping [34] or Longest Common Subsequence measures [11]. The general idea is to limit the elasticity of the measure by using a threshold: if the index difference between two samples that are candidates for a match is lower than the value of the threshold, then matching is allowed, otherwise it is forbidden. This binary decision might, in some cases, limit the effectiveness of the measure. Keeping in mind the mechanical analogy of a spring (for which the deformation effort is proportional to the stretching or the shrinking), instead of using a threshold we suggest using the range of the sample index difference to linearly penalize the matching of samples for which the index values are too far and to favor the matching of samples for which the index values are close. In the case where time series are sampled using non uniform or varying sampling rates one can benefit from time stamps instead of sample indices since this approach does not require resampling of the data.

The second motivation for defining a measure that exploits time stamps (or sample indices) while verifying the triangle inequality is two folds: first, it provides an effective solution for comparing approximated representations of time series, but not necessarily by using uniform down-sampling methods; second, it establishes a useful relationship between the matching performed in the down-sampled space and the matching performed in the original space.

Approximation of multi dimensional discrete curves has been widely studied [7][13][31] essentially to speed up the data processing required by resource demanding applications. Among other approaches, polygonal approximation of discrete curves has been quite popular recently [31][17]. The problem can be informally stated as follows: given a digitized curve $X$ of $N \geq 2$ ordered samples, find $K$ (in general $K<<N$) dominant samples among them that define a sequence of piecewise linear segments which most closely approximate the original curve. This problem is known as the *min-ε* problem [12]. Numerous algorithms have been proposed for over thirty years to solve this optimization problem efficiently. Most of them belong either to graph-theoretic, dynamic programming or to heuristic approaches. See for instance [13] [31] [17] [24] among others for details.

Such approaches can be used to adaptively down sample time series. For instance, in [24] polygonal curves approximations have been used to down sample gesture signals optimally and in [24] an elastic matching procedure has been proposed to compare two time series with a linear time complexity. For these approaches, a down sampled time series is a reduced sequence of tuples (sample, time stamps) that corresponds to the end extremities of the polygonal segments. The sampling rate for such down-sampled time series is not generally uniform in.

Down sampling time series can be used to drastically reduce the dimension of the space in which we could potentially process the time series. Nevertheless, one difficulty emerges: how can we compare down-sampled time series using non uniform (e.g. varying) sampling frequencies? Not taking into account the occurring time of the samples could introduce discrepancies between the original space and the down-sampled space. For instance, phase or frequency information is potentially lost or at least damaged, as well as the slope of spikes.

In this context, the triangle inequality is also of great importance since it maintains distance relations between the original space and the down-sampled space. Let $X$ and $Y$ be two time series in the original space and $\tilde{X}$ and $\tilde{Y}$ their down-sampled counter parts. If $\delta$ is a measure for which the triangle inequality holds, then we have: $|\delta(X,Y) - \delta(\tilde{X},\tilde{Y})| \leq \delta(X,\tilde{X}) + \delta(Y,\tilde{Y})$

In the case where $\delta(X,\tilde{X})$ and $\delta(Y,\tilde{Y})$ are maintained small by the similarity measure, $\delta(X,Y)$ and $\delta(\tilde{X},\tilde{Y})$ are comparable and the following inequality gives an exploitable lower bound to the $\delta(X,Y)$ measure:

$$\delta(\tilde{X},\tilde{Y}) - \delta(X,\tilde{X}) - \delta(Y,\tilde{Y}) \leq \delta(X,Y) \quad (1)$$

This lower bound can be used to significantly speed up the time series information retrieval process since a pruning strategy can be proposed in the down-sampled space. We will come back to this issue in section V.

III. ELASTIC SIMILARITY IN LIGHT OF THE SYMBOLIC EDIT DISTANCE

In this section we succinctly present the main elastic measures developed in the literature, from founding work to more recent studies.

The Levenshtein Distance (LD) proposed in 1966 [19], also known as the edit distance, is the smallest number of insertions, deletions, and substitutions required to change one string into another. For more than thirty years, the ideas behind LD have been largely reused and extended by various research communities. The main contributions are rapidly reviewed below. In 1974 Wagner and Fisher [38] developed a computationally efficient algorithm to calculate LD in O(n.m) using dynamic programming [2]. Meanwhile Dynamic Time Warping, which shares many similarities with LD despite the fact that it is not a metric, was proposed in 1970 [36] and 1971 [34] to align speech utterances, namely time series, with time shift tolerances. The Longest Common Subsequence (LCSS) similarity measure initially defined for string matching [11] has also been adapted for time series matching [6][37]. Recently, a lot of fruitful research dealing with DTW and LCSS has been carried out to propose efficient computation and pruning strategies that are required to process massive data [37][14][40]. Some work has also been conducted to provide the 'triangle inequality' to DTW: the Edit Distance with Real Penalty (ERP) [5] has been proposed as an edit distance based metric for time series matching with time shift tolerance. The edit distance principle has also been proposed to develop 1D-Point-Patterns Matching (PPM) (point patterns are ascending lists of real values) [21][22]. The measure proposed to match 1D-Point Patterns is shown to be a metric that can be extended to the multidimensional case, at the price of a non polynomial complexity. Hereinafter, we present DTW, ERP and LCSS in light of the edit distance and develop the TWED metrics as an alternative to ERP.

*A. Definitions*

Let $U$ be the set of finite time series: $U = \{A_1^p \,/\, p \in N\}$. $A_1^p$ is a time series with discrete time index varying between $1$ and $p$. We note $\Omega$ the empty time series (with null length) and by convention $A_1^0 = \Omega$ so that $\Omega$ is member of set $U$.

Let $A$ be a finite discrete time series. Let $a'_i$ be the $i^{th}$ sample of time series $A$. We will consider that $a'_i \in S \times T$ where $S \subset R^d$ with $d \geq 1$ embeds the multidimensional space variables and $T \subset R$ embeds the time stamp variable, so that we can write $a'_i = (a_i, t_{a_i})$ where $a_i \in S$ and $t_{a_i} \in T$, with the condition that $t_{a_i} > t_{a_j}$ whenever $i > j$ (time stamp strictly increase in the sequence of samples).

$A_i^j$ with $i \leq j$ is the sub time series consisting of the $i^{th}$ through the $j^{th}$ samples (inclusive) of $A$. So $A_i^j = a'_i \, a'_{i+1} \ldots a'_j$. $|A|$ denotes the length (the number of samples) of $A$. $\Lambda$ denotes the null sample. $A_i^j$ with $i > j$ is the null time series noted $\Omega$.

An edit operation is a pair $(a', b') \neq (\Lambda, \Lambda)$ of time series samples, written $a' \to b'$. Time series $B$ results from the application of the edit operation $a \to b$ into time series $A$, written $A \Rightarrow B$ *via* $a' \to b'$, if $A = \sigma a' \tau$ and $B = \sigma b' \tau$ for some time series $\sigma$ and $\tau$. We call $a' \to b'$ a match operation if $a' \neq \Lambda$ and $b' \neq \Lambda$, a delete operation if $b' = \Lambda$, an insert operation if $a' = \Lambda$.

Similarly to the edit distance defined for string [11], we define $\delta(A, B)$ as the similarity between any two time series $A$ and $B$ of finite length, respectively $p$ and $q$ as:

$$\delta(A_1^p, B_1^q) = Min \begin{cases} \delta(A_1^{p-1}, B_1^q) + \Gamma(a'_p \to \Lambda) & delete \\ \delta(A_1^{p-1}, B_1^{q-1}) + \Gamma(a'_p \to b'_q) & match \\ \delta(A_1^p, B_1^{q-1}) + \Gamma(\Lambda \to b'_q) & insert \end{cases} \quad (2)$$

Where $p \geq 1$, $q \geq 1$ and $\Gamma$ is an arbitrary cost function which assigns a nonnegative real number $\Gamma(a' \to b')$ to each edit operation $a' \to b'$.

The recursion is initialized by setting:

$$\delta(A_1^0, B_1^0) = 0$$
$$\delta(A_1^0, B_1^j) = \infty$$
$$\delta(A_1^i, B_1^0) = \infty$$

Dynamic Time Warping (DTW) and Edit Distance with Real penalties (ERP), 1D Point Pattern matching (PPM) and Longest Common Subsequence (LCSS) are special cases of the previous definitions:

*B. The DTW special case*

The DTW similarity measure [36][34] $\delta_{DTW}$ is defined according to the previous notations as:

$$\delta_{DTW}(A_1^p, B_1^q) = d_{LP}(a_p, b_q) + Min \begin{cases} \delta_{DTW}(A_1^{p-1}, B_1^q) \\ \delta_{DTW}(A_1^{p-1}, B_1^{q-1}) \\ \delta_{DTW}(A_1^p, B_1^{q-1}) \end{cases} \quad (3)$$

where $d_{LP}(x, y)$ is the $Lp$ norm of vector $x$-$y$ in $R^d$,

and so for DTW, $\Gamma(a'_p \to \Lambda_q) = \Gamma(\Lambda_p \to b'_q) = \Gamma(a'_p \to b'_q) = d_{LP}(a_p, b_q)$

One may note that the time stamp values are not used, therefore the costs of each edit operation involve vectors $a$ and $b$ in $S$ instead of vectors $a'$ and $b'$ in $S \times T$. One of the main restrictions of $\delta_{DTW}$ is that it does not comply with the triangle inequality as shown by the following example [5]:

$$A_1^1 = [1]; \; B_1^2 = [1,2]; \; C_1^3 = [1,2,2](*)$$
$$\delta_{DTW}(A_1^1, B_1^2) = 1; \; \delta_{DTW}(B_1^2, C_1^3) = 0; \; \delta_{DTW}(A_1^1, C_1^3) = 2$$
$$\text{and thus}: \delta_{DTW}(A_1^1, C_1^3) > \delta_{DTW}(A_1^1, B_1^2) + \delta_{DTW}(B_1^2, C_1^3)$$

(*) *1D time series with no stamp value given*

*C. The ERP special case*

$$\delta_{ERP}(A_1^p, B_1^q) = Min \begin{cases} \delta_{ERP}(A_1^{p-1}, B_1^q) + \Gamma(a'_p \to \Lambda) \\ \delta_{ERP}(A_1^{p-1}, B_1^{q-1}) + \Gamma(a'_p \to b'_q) \\ \delta_{ERP}(A_1^p, B_1^{q-1}) + \Gamma(\Lambda \to b'_q) \end{cases} \quad (4)$$

$$\text{with} \quad \begin{aligned} \Gamma(a'_p \to \Lambda) &= d_{LP}(a_p, g) \\ \Gamma(a'_p \to b'_q) &= d_{LP}(a_p, b_q) \\ \Gamma(\Lambda \to b'_q) &= d_{LP}(g, b_q) \end{aligned}$$

and *g* a constant in *S*.

where $d_{LP}(x, y)$ is the *Lp* norm of vector *x-y* in *S*. Note that the time stamp coordinate is not taken into account, therefore $\delta_{ERP}$ is a distance on S but not on $S \times T$.

Here again, it can be noted that the time stamp values are not used, thus the costs of each edit operation involve vectors *a* and *b* in $R^d$ instead of vectors *a'* and *b'* in $R^{d+1}$.

According to the authors of ERP [5], the constant *g* should be set to *0* for some intuitive geometric interpretation and in order to preserve the mean value of the transformed time series when adding gap samples.

*D. The LCSS special case*

The Longest Common Subsequence (LCSS) similarity measure has been first defined for string matching purposes [11] and then extended for times series [6][37]. LCSS is recursively defined in [37] as follows:

$$LCSS_{\varepsilon,\delta}(A_1^p, B_1^q) = \begin{cases} 0 \text{ if } p < 1 \text{ or } q < 1, \\ 1 + LCSS_{\varepsilon,\delta}(A_1^{p-1}, B_1^{q-1}) \text{ if } d_{LP}(a_p, b_q) < \varepsilon \text{ and } |p - q| < \delta, \\ Max\{LCSS_{\varepsilon,\delta}(A_1^{p-1}, B_1^q), LCSS_{\varepsilon,\delta}(A_1^p, B_1^{q-1})\} \text{ otherwise} \end{cases} \quad (5)$$

For LCSS the match reward is *1*, while no reward is offered for *insert* or *delete* operations.

The LCSS measure is transposed into a normalized dissimilarity measure $D_{\varepsilon,\delta}$ which is close in its formal structure to the ERP measure:

$$D_{\varepsilon,\delta} = 1 - \frac{LCSS_{\varepsilon,\delta}(A_1^p, B_1^q)}{Min\{p,q\}} \quad (6)$$

### E. The 1D PPM special case

For Point-Pattern matching problems [21], $A_1^p$ and $B_1^q$ are *1D* ascending lists of real values.

$$\delta_{PPM}(A_1^p, B_1^q) = Min \begin{cases} \delta_{PPM}(A_1^{p-1}, B_1^q) + \Gamma(a'_p \to \Lambda) \\ \delta_{PPM}(A_1^{p-1}, B_1^{q-1}) + \Gamma(a'_p \to b'_q) \\ \delta_{PPM}(A_1^p, B_1^{q-1}) + \Gamma(\Lambda \to b'_q) \end{cases} \quad (7)$$

$$\text{with} \quad \begin{aligned} \Gamma(a'_p \to \Lambda) &= a_p - a_{p-1} \\ \Gamma(a'_p \to b'_q) &= |(a_p - a_{p-1}) - (b_q - b_{q-1})| \quad \text{if } p>1 \text{ and } q>1 \\ \Gamma(\Lambda \to b'_q) &= b_q - b_{q-1} \end{aligned}$$

The author [21] shows that $\delta_{PPM}$ is a metric that calculates the minimum amount of space needed to delete or insert between pairs of points to convert one point-pattern into another. It can be noted that if successive increments are considered instead of the initial values, $\delta_{PPM}$ coincides with the $\delta_{ERP}$ applied to the lists of positive increments.

### F. Symbolic sequence alignment with affine gap penalty

In biomolecular sequences (DNA, RNA, or amino acid sequences), high sequence similarity usually implies significant functional or structural similarity. The basic mutational processes behind the evolution of such sequences are substitutions, insertions and deletions, the latter two giving rise to gaps. Various similarity models based on dynamic programming have been developed by the bioinformatics community. Among them, the affine gap model [10] [8] that extends the Needleman-Wunsch algorithm [28] should be mentioned. The originality of this model is to penalized gap sequences according to the affine equation $\gamma(g) = -d - (g-1).e$, where $g$ is the length of the gap, $d$ is the *open-gap* penalty and $e$ is the *gap-extension* penalty. The recursion is given in equation (8):

$$M(A_1^p, B_1^q) = \Gamma(a_p \rightarrow b'_q) + Max \begin{cases} I_x(A_1^{p-1}, B_1^{q-1}) \\ M(A_1^{p-1}, B_1^{q-1}) \\ I_y(A_1^{p-1}, B_1^{q-1}) \end{cases}$$

$$I_x(A_1^p, B_1^q) = Max \begin{cases} M(A_1^{p-1}, B_1^q) - d \\ I_x(A_1^{p-1}, B_1^q) - e \end{cases} \quad (8)$$

$$I_x(A_1^p, B_1^q) = Max \begin{cases} M(A_1^p, B_1^{q-1}) - d \\ I_x(A_1^p, B_1^{q-1}) - e \end{cases}$$

Here $\Gamma(a_p \rightarrow b_q)$ is an integer value either positive when $a_p$ and $b_q$ are similar symbols, or negative when $a_p$ and $b_q$ are dissimilar symbols, $M(A_1^p, B_1^q)$ is the best score up to $(a_p, b_q)$, given that $a_p$ is aligned to $b_q$, $I_x(A_1^p, B_1^q)$ is the best score up to $(a_p, b_q)$, given that $a_p$ is aligned to a gap, and $I_x(A_1^p, B_1^q)$ is the best score up to $(a_p, b_q)$, given that $b_q$ is aligned to a gap. The previous recursions are initialized as follows:

$$M(A_1^0, B_1^0) = 0, \quad I_x(A_1^0, B_1^0) = I_y(A_1^0, B_1^0) = -\infty$$
$$M(A_1^i, B_1^0) = I_x(A_1^i, B_1^0) = -d - (i-1).e, \quad I_y(A_1^i, B_1^0) = -\infty, \text{ for } i = 1,...p$$
$$M(A_1^0, B_1^j) = I_y(A_1^0, B_1^j) = -d - (j-1).e, \quad I_x(A_1^0, B_1^j) = -\infty, \text{ for } j = 1,...q$$

IV. THE TWED DISTANCE

We propose an alternative way of defining of the edit operations for time series alignment which leads to the definition of the new similarity measure TWED. To understand the semantic associated to the edit operations for TWED, we reconsider the editing analogy with strings and suggest some differences. The edit distance between two strings is defined as the minimal transformation cost allowing for the transformation of the first string into the second one. For string edition, a transformation is a finite sequence of edit operations whose associated cost is the sum over the sequence of edit operations of the elementary costs $\Gamma$ associated to each edit operation.

### A. Graphical Editor Paradigm

For discrete time series we are seeking a sequence of edit operations allowing for the simultaneous transformation of two time series to superimpose them with a minimal cost. If we use a graphical editor paradigm, we can imagine a *2D* representation of time series for which the horizontal axis represents the time scale or the time stamp coordinate and the vertical axis represents a spatial coordinate scale displaying the projection of the *d-1* spatial coordinates of the samples onto a *1D* scale. In this display, discrete time series are considered as a sequence of linear segments between successive samples. The graphical editor we have imagined allows for the editing of two time series *A* and *B* using three elementary edit operations depicted in Fig. *1a, 1.b* and *1.c*.

Instead of the classical *delete*, *insert* and *match* operations, we introduce *delete-A*, *delete-B* and *match* operations as follows:

i) The *delete-A* (delete inside the first time series) operation (Fig. *1.b*) consists of clicking on the dot which represents the sample in *A* to delete ($a'_i$) and of dragging and dropping this dot onto the previous sample dot ($a'_{i-1}$). We suggest that the editing effort or cost associated with this delete operation is proportional to the length of vector ($a'_i - a'_{i-1}$) to which we add a constant penalty $\lambda \geq 0$.

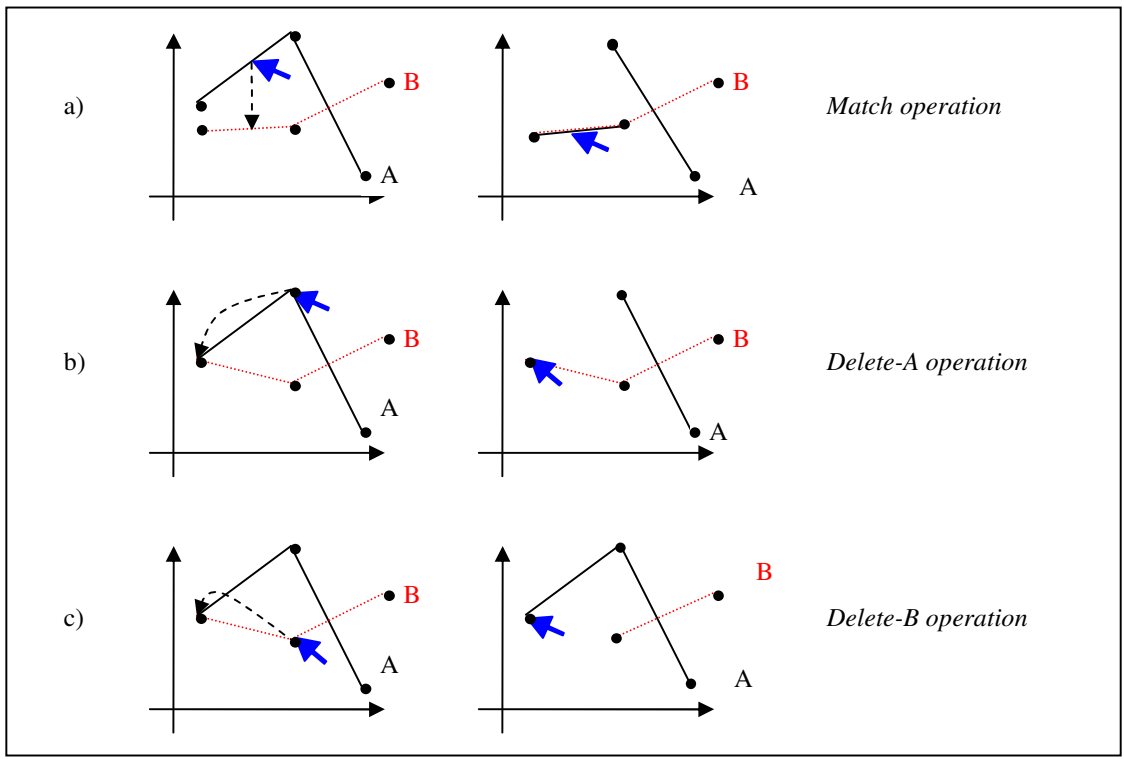

*Fig 1: The edit operations in the graphical editor paradigm.*

ii) The *delete-B* (delete inside the second time series) operation (Fig. *1.c*) consists of clicking on the dot which represents the sample in *B* to delete ($b'_i$) and of dragging and dropping this dot onto the previous sample dot ($b'_{i-1}$). Here again, we suggest that the editing effort or cost associated with this delete operation is proportional to the length of vector ($b'_i - b'_{i-1}$) to which we add a constant penalty $\lambda \geq 0$.

Due to sampling rate variations or process variability one could be faced the situation where in time series data, one event could be registered many times or only few times when recording different utterances; this would justify that the deletion cost be proportional to the distance to the previous sample. Nevertheless 'outlier' samples (e.g. spurious data points) deletion cannot be covered by this argument. According to TWED, the deletion cost for such sample depends on the previous sample in the time series, and there is no specific argument to justify it. The other elastic measures (DTW, ERP, LCSS) do not offer better justification for the deletion cost of 'outliers'.

iii) The *match* operation (Fig. *1*.*a*) consists of clicking on the segment to match in the first time series ( $a'_{i-1}$ $a'_i$ ) and then of dragging and dropping this segment onto the graphic position corresponding to the matching segment ($b'_{j-1}$ $b'_j$ ) in the second time series. We can suggest that the editing effort or cost associated with the *match* operation is proportional to the sum of the lengths of the two vectors ($b'_j - a'_i$) and ($b'_{j-1} - a'_{i-1}$).

This provides the basis for the TWED distance we propose

## B. Definition of TWED

$$\delta_{\lambda,\gamma}(A_1^p, B_1^q) = Min \begin{cases} \delta_{\lambda,\gamma}(A_1^{p-1}, B_1^q) + d(a'_p, a'_{p-1}) + \lambda & delete - A \\ \delta_{\lambda,\gamma}(A_1^{p-1}, B_1^{q-1}) + d(a'_p, b'_q) + d(a'_{p-1}, b'_{q-1}) & match \\ \delta_{\lambda,\gamma}(A_1^p, B_1^{q-1}) + d(b'_{q-1}, b'_q) + \lambda & delete - B \end{cases} \quad (9)$$

The recursion is initialized setting:
$$\delta_{TWED}(A_1^0, B_1^0) = 0$$
$$\delta_{TWED}(A_1^0, B_1^j) = \infty$$
$$\delta_{TWED}(A_1^i, B_1^0) = \infty$$
with $a'_0 = b'_0 = 0$    by convention.

It is interesting to note that the penalties for *delete-A* or *delete-B* operations are similar to those proposed in the $\delta_{PPM}$ measure if we do not consider the time stamps coordinate and address the matching of *1D* monotone increasing time series.

Furthermore, using the graphical editor paradigm, we define the time series matching game as follows: two time series, *A* and *B*, are displayed on the graphic. The goal is to edit *A* and *B* to completely superimpose the two curves.

The editing process is performed from left to right: if $i$ is an index on the segments of $A$ and $j$ on the segments of $B$, then the process's initial setting is $i=j=1$. A match operation will increment $i$ and $j$ simultaneously: $i \leftarrow i+1$, $j \leftarrow j+1$. A *delete-A* operation will increment $i$ only: $i \leftarrow i+1$. A *delete-B* operation will increment $j$ only: $j \leftarrow j+1$.

According to the above mentioned constraint, once a segment $i$ in $A$ has been processed using either a *match* or a *delete-A* operation, it is impossible to edit it again: this rule applies for all previous segments $r$ in $\{1,...,i-1\}$. Similarly, once a segment $j$ in $B$ has been used either in a *match* or in an *delete-B* operation it is impossible to use former samples $r$ in $\{1..j\}$ for future match or insertion operations. Therefore, according to this game, the editing process provides a sequence of edit operations as well as ordered pairs of indices $(i,j)$ where $i$ is an index in the sequence of segments of $A$ and $j$ an index in the sequence of segments of $B$. In other words, the process provides an ordered sequence of triplets $(op_k, i_k, j_k)$ where $op_k$ is the $k^{th}$ edit operation selected, and $i_k$ and $j_k$ are the values of the index in $A$ and $B$ respectively when the edit operation is performed. A partial order can be defined on the triplets as follows:

$(op_{k1}, i_{k1}, j_{k1}) < (op_{k2}, i_{k2}, j_{k2})$ iff $i_{k1} \leq i_{k2}$ and $j_{k1} \leq j_{k2}$ and either $i_{k1} \neq i_{k2}$ or $j_{k1} \neq j_{k2}$.

Since for each step of the editing game, one of the indices is increased by one while the other is either incremented by one or remains unchanged, all the triplets in the output editing sequence are ordered in increasing order.

Supposing that the game editing process has provided a sequence of edit operations up to $i_k$ and $j_k$ index values, if the sub sequences $\overline{A}_1^{i_{k-1}}$ ($\overline{A}_1^{i_{k-1}}$ refers to the sequence obtained from $A$ after the first $k$-$1$ edit operations) and $\overline{B}_1^{j_{k-1}}$ are not superimposed, then, as there is no possibility to process former samples so that they may be superimposed, the game process cannot be successful.

It is easy to show that $\delta_{\lambda,\gamma}$ as defined in eq. (9) provides a successful sequence of editing operations at a minimal global cost for all pairs of time series in $U^2$.

### C. Some properties of TWED

**Proposition 1:** $\delta_{\lambda,\gamma}$ is a distance on the set of finite discrete time series $U$:

P1: $\delta_{\lambda,\gamma}(A,B) \geq 0$ for any finite discrete time series $A$ and $B$,

P2: $\delta_{\lambda,\gamma}(A,B) = 0$ iff $A = B$ for any finite discrete time series $A$ and $B$,

P3: $\delta_{\lambda,\gamma}(A,B) = \delta_{\lambda,\gamma}(B,A)$ for any finite discrete time series $A$ and $B$,

P4: $\delta_{\lambda,\gamma}(A,B) \leq \delta_{\lambda,\gamma}(A,C) + \delta_{\lambda,\gamma}(C,B)$ for any finite discrete time series $A$, $B$ and $C$.

Proof of Proposition 1 is given in [25].

**Proposition 2:** $\delta_{\lambda,\gamma}$ is upper bounded by twice the distance $D_{LP}$.

$\forall \lambda \geq 0, \gamma > 0 \quad \forall X, Y \in U^2, \quad \delta_{\lambda,\gamma}(X,Y) \leq 2 \cdot D_{LP}(X,Y)$, whenever $X$ and $Y$ have the same length. The proof is given in [25].

**Proposition 3:** $\delta_{\lambda,\gamma}$ is an increasing function of $\lambda$ and $\gamma$:

$$\forall \lambda \geq 0, \gamma > 0 \; \forall \lambda' \geq \lambda \; \forall \gamma' \geq \gamma \; \forall X,Y \in U^2 \quad \delta_{\lambda,\gamma}(X,Y) \leq \delta_{\lambda',\gamma'}(X,Y)$$

The proof is given in [25].

### D. Providing 'stiffness' into $\delta_{\lambda,\gamma}$

Going back to the graphical editor game we have envisaged that the penalty or cost associated with each edit operation should be proportional to the mouse pointer displacement involved during the edition. If we separate the spatial displacement in $S$ from the temporal displacement in $T$ then we have to consider a spatial penalty that could be handled by a distance measured in $S$ and a temporal penalty more or less proportional to some distance measured in $T$. By doing so, we could parameterize a distance in between the Euclidian Distance, which is characterized by a kind of 'infinite stiffness', and DTW which is characterized by a 'null stiffness. In practice, we can choose $d(a',b') = d_{LP}(a,b) + \gamma \cdot d_{Lp}(t_a,t_b)$ where $\gamma$ is a non negative constant which characterizes the *stiffness* of $\delta_{\lambda,\gamma}$ elastic measures. Notice that $\gamma > 0$ is required for $\delta_{\lambda,\gamma}$ to be a distance. If $\gamma = 0$ then $\delta_{\lambda,\gamma}$ will be a distance on $S$ but not on $S \times T$.

The final formulation of $\delta_{\lambda,\gamma}$ is as follows:

$$\delta_{\lambda,\gamma}(A_1^p, B_1^q) = Min \begin{cases} \delta_{\lambda,\gamma}(A_1^{p-1}, B_1^q) + d_{LP}(a_p, a_{p-1}) + \gamma \cdot d_{Lp}(t_{a_p}, t_{a_{p-1}}) + \lambda \\ \delta_{\lambda,\gamma}(A_1^{p-1}, B_1^{q-1}) + d_{LP}(a_p, b_q) + \gamma \cdot d_{Lp}(t_{a_p}, t_{b_q}) + d_{LP}(a_{p-1}, b_{q-1}) + \gamma \cdot d_{Lp}(t_{a_{p-1}}, t_{b_{q-1}}) \\ \delta_{\lambda,\gamma}(A_1^p, B_1^{q-1}) + d_{LP}(b_q, b_{q-1}) + \gamma \cdot d_{Lp}(t_{b_q}, t_{b_{q-1}}) + \lambda \end{cases} \quad (10)$$

Some analogy can be found between the parameters $\lambda$ and $\gamma$ of $\delta_{\lambda,\gamma}$ and parameters $e$ and $d$ of the affine model defined for symbolic sequence matching (see section III.F). Nevertheless, some major differences exist: the penalties in $\delta_{\lambda,\gamma}$ are, for one part, proportional to the time stamp difference between matching, deleted or inserted samples. A constant penalty is added for the two deletion operations that correspond to gaps. Conversely, the affine model proposes a penalty proportional to the gap length corresponding to series of successive insertions or deletions, with a constant penalty for the first operation in the sequence.

The iterative implementation of $\delta_{\lambda,\gamma}$ using the *Lp* metrics to evaluate the distance between two samples is depicted in *Fig.2*.

### E. Algorithmic complexity of $\delta_{\lambda,\gamma}$

The time complexity of $\delta_{\lambda,\gamma}$ is the same as DTW and ERP, namely *O(p.q)*, where *p* and *q* are the lengths of the two time series being matched. The space complexity is also the same as DTW i.e. *O(p.q)*, but as the ERP distance as well, the costs $\Gamma(a' \rightarrow \Lambda)$ and $\Gamma(\Lambda \rightarrow b')$ can be tabulated to speed up the calculation leading to an extra space complexity of *O(p+q)* for $\delta_{TWED}$.

```
float TWED(float A[1..n], float timeStampsA[1..n],
          float B[1..m], float timeStampsB[1..m],
          float lambda, float nu) {
   declare int DTW[0..n,0..m];
   declare int i, j;
   declare float cost;
   declare float A[0] :=0, timeStampsA[0] :=0;
   declare float B[0] :=0, timeStampsB[0] :=0;

   for i := 1 to m
       TWED[0,i] := infinity;
   for i := 1 to n
       TWED[i,0] := infinity;
   TWED[0,0] := 0;

   for i := 1 to n {
       for j := 1 to m {
           cost:= LpDist(A[i],B[j]);   // Distance-L1
           DTW[i,j] := minimum(
               // insertion
               DTW[i-1,j  ] + LpDist(A[i-1], A[i])+
                      nu*(timeStampsA[i]- timeStampsA[i-1]+lambda,
               // deletion
               DTW[i  ,j-1] + LpDist(B[j-1], B[j])+
                      nu*(timeStampsB[j]- timeStampsB[j-1]+lambda,
               // match
               DTW[i-1,j-1] + LpDist(A[i],B[j])+
                      nu*|timeStampsA[i]- timeStampsB[j]| )+
                      LpDist(A[i-1],B[j-1])+
                      nu*|timeStampsA[i-1]- timeStampsB[j-1]| );
       } // End for j
   } // End for i

   Cost = TWED[n,m];
   Return; }
```

*Fig 2: Iterative implementation of the TWED distance.*

## V. BOUNDING THE TWED MEASURE

In this section we get back to our second motivation about defining a measure that exploits time stamps while verifying the triangle inequality. We show how piecewise constant approximations (PWCA) with few segments of time series can be used to improve the efficiency of range queries. Various methods exist to get polygonal curve approximations of time series, in particular heuristic [7][9][13][15], near optimal [17][24] or optimal [31] solutions. Most of them can be adapted to provide PCWA approximation of time series.

We define $\overline{A}_1^{p,r}$ as a PWCA of time series $A_1^p$ containing $r-1 \geq 0$ constant segments and $p$ samples. This approximation can be obtained using any kind of solution (from heuristic to optimal solutions), let say the optimal solution similar to the one proposed in [31]. $\overline{A}_1^{p,r}$ and $A_1^p$ have the same number of samples, namely $p$. Let $\tilde{A}_1^r$ be the time series composed with the $r$ segment extremities of $\overline{A}_1^{p,r}$. $\tilde{A}_1^r$ contains $r$ samples. Let us similarly define $\overline{B}_1^{p,r'}$ and $\tilde{B}_1^{r'}$ from time series $B_1^p$.

***Proposition 4:***

$\forall \lambda \geq 0, \gamma > 0, \quad \forall r \in [1; p[, \quad \forall X_1^p \in U \quad \delta_{\lambda,\gamma}(\overline{X}_1^{p,r}, \tilde{X}_1^r) \leq \lambda \cdot (p-r) + \gamma \cdot \Delta T(2 \cdot p - r)$, where $\Delta T$ is the time difference average between two successive samples inside the piecewise constant segments of the approximation.

The proof of this proposition is given in [25].

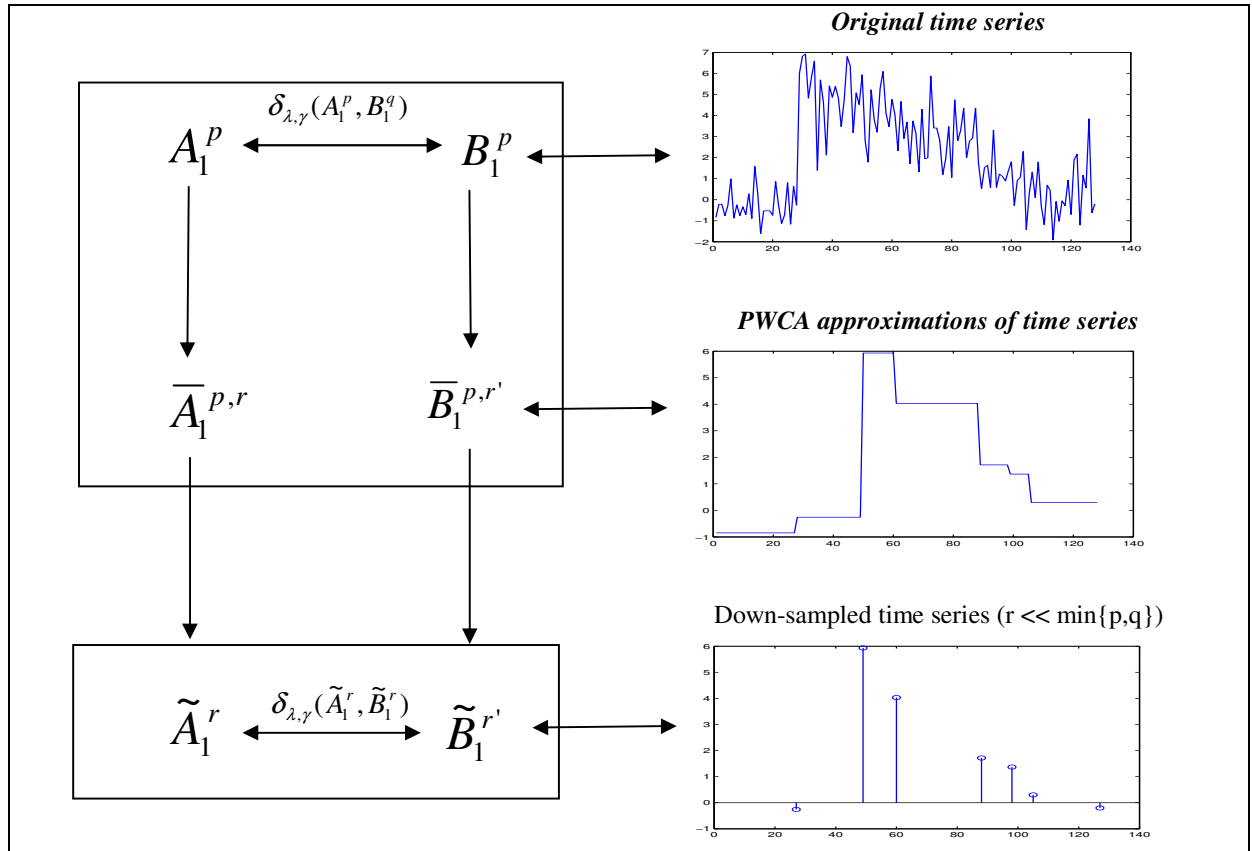

*Fig 2: Linking the matching of time series in the original space to the matching in the down-sampled space.*

From these previous propositions and the triangular inequality property we get an upper-bound for $\left|\delta_{\lambda,\gamma}(A_1^p, B_1^q) - \delta_{\lambda,\gamma}(\tilde{A}_1^r, \tilde{B}_1^{r'})\right|$ that quantifies the difference of the distance of two time series evaluated in the original space with the distance of their approximations evaluated in the down-sampled (see appendix for details):

$$\left|\delta_{\lambda,\gamma}(A_1^p, B_1^q) - \delta_{\lambda,\gamma}(\tilde{A}_1^r, \tilde{B}_1^{r'})\right| \leq \lambda \cdot (p + q - r - r') + \gamma \cdot \Delta T(2 \cdot (p+q) - r - r') + 2 \cdot D_{LP}(\overline{A}_1^{p,r}, A_1^p) + 2 \cdot D_{LP}(\overline{B}_1^{q,r'}, B_1^q)$$

This shows that $\delta_{\lambda,\gamma}(A_1^p, B_1^q)$ and $\delta_{\lambda,\gamma}(\tilde{A}_1^r, \tilde{B}_1^{r'})$ are potentially close when two conditions are satisfied:

1. The PWCA approximations of *A* and *B* are close to the original time series in the sense of the *LP-distance*. This should be ensured by the optimal solution of the *min-ε* problem using piecewise constant segments whenever the number of segments *r* is not too small.

2. $\lambda$ and $\gamma \cdot \Delta T$ are small comparatively to $2 \cdot (p+q) - r - r'$

Hence, we get the following lower bounds that can be considered tight if the two previous conditions are satisfied:

$$\forall \lambda \geq 0, \gamma > 0, \ \forall \lambda' \geq \lambda \ \forall \gamma' \geq \gamma, \ \forall A_1^p, B_1^q \in U^2, \ \forall r \in [1, p[, \forall r' \in [1, q[,$$
$$\delta_{\lambda,\gamma}(\tilde{A}_1^r, \tilde{B}_1^{r'}) - \lambda \cdot (p+q-r-r') + \gamma \cdot \Delta T(2 \cdot (p+q) - r - r') - 2 \cdot D_{LP}(\overline{A}_1^{p,r}, A_1^p) - 2 \cdot D_{LP}(\overline{B}_1^{q,r'}, B_1^q)$$
$$\leq \delta_{\lambda,\gamma}(\tilde{A}_1^r, \tilde{B}_1^{r'}) - \delta_{\lambda,\gamma}(\tilde{A}_1^r, A_1^p) - \delta_{\lambda,\gamma}(\tilde{B}_1^{r'}, B_1^p) \ \leq \delta_{\lambda,\gamma}(A_1^p, B_1^q) \leq \delta_{\lambda',\gamma'}(A_1^p, B_1^q) \quad (12)$$

This last inequality is still potentially useful to design fast and dirty filters dedicated to range query searching for applications for which $\lambda'$ and $\gamma'$ cannot be small enough, while $\lambda$ and $\gamma$ can be set up small. For range query search, if *R* is the radius of the range query and $A_1^p$ the center of the query ball, then $B_1^q$ is outside the search range if one of the following conditions is verified:

$$\delta_{\lambda,\gamma}(\tilde{A}_1^r, \tilde{B}_1^{r'}) > R + \lambda \cdot (p+q-r-r') + \gamma \cdot \Delta T(2 \cdot (p+q) - r - r') + 2 \cdot D_{LP}(\overline{A}_1^{p,r}, A_1^p) + 2 \cdot D_{LP}(\overline{B}_1^{q,r'}, B_1^q)$$
$$> R + \delta_{\lambda,\gamma}(\tilde{A}_1^r, A_1^p) + \delta_{\lambda,\gamma}(\tilde{B}_1^{r'}, B_1^p) \quad (13)$$

For time series information retrieval applications, inequalities (12) and (13) are potentially useful. If $\delta_{\lambda,\gamma}(\tilde{A}_1^r, A_1^p)$ and $\delta_{\lambda,\gamma}(\tilde{B}_1^{r'}, B_1^p)$ are pre-computed during the indexing phase, the tighter bound can be used. Otherwise the second bound can be evaluated during the retrieval phase through the computation of *L1-distances*.

If $r = r' = (1/K.)\min(p,q)$ the complexity for evaluating $\delta_{\lambda,\gamma}(\widetilde{A}_1^r, \widetilde{B}_1^r)$ is lower than $O(p.q/K^2)$.

## VI. EXPERIMENTATIONS

### A. Classification task experiment

To evaluate empirically the effectiveness of the TWED distance comparatively to other metrics or similarity measures, we address a simple classification task experiment. The classification task we have considered consists of assigning one of the possible categories to an unknown time series for the 20 data sets available at UCR repository [16].

For each dataset, a training subset is defined as well as a testing subset. The classification is based on the simple nearest neighbor decision rule: first we select a training data set containing time series for which the correct category is known. To assign a category to an unknown time series selected from a testing data set (different from the train set), we select its nearest neighbor (in the sense of a distance or similarity measure) within the training data set, then, assign the associated category to its nearest neighbor.

Given a dataset, we adapt the *stiffness* parameter as follows: we use the **training dataset** to select the '*best stiffness*' ($\gamma$) value as well as the best $\lambda$ value, namely the ones leading to the minimal error rate **on the training data**, according to a leave-one-out procedure (that consists of iteratively selecting one time series from the **training set** and then in considering it as a test against the remaining time series within the **training set** itself).

Finally, the **testing dataset** is used to evaluate the final error rate (reported in Tab.1 and Tab.2) with the best $\gamma$ and $\lambda$ values estimated on the **training set**. This leads to OTWED, the optimized versions of TWED. The same procedure is used to set up the parameters defined for the other parametric measures, i.e. ODTW and LCSS.

*Tab.1* and *Tab.2* show the results obtained for the tested methods, e.g. Euclidian Distance on the original time series, optimized DTW with best warping windows (ODTW) as defined in [32], classical DTW (DTW) with no warping window, Longest Common Subsequence (LCSS) as defined in [37], Edit distance with Real Penalty *(ERP)* as defined in [5] and OTWED. In *Tab.1* and *Fig.3* the time series are not preprocessed, while in *Tab.2* and *Fig 4.* time series are down sampled using an optimal Piecewise Constant Approximation procedure similar as the one described in [31] for polygonal approximation. In this last experiment, each down sampled time series has exactly 50% less samples than the original time series. The sampling rate for the down sampled time series is indeed varies, since the size of the constant segments used to approximate the time series is not generally constant.

For parameterized measures, best values are selected from the training data in order to minimize the error rate estimated for the training data. More precisely the settings are as follows:

- ODTW: the best corridor value is selected for each dataset among the set *{0, max{p,q}}* so as to minimize the classification errors estimated for the **training data**. If different corridor values lead to the minimal error rate estimated for the training data then the **lowest corridor value is selected.**

- LCSS: the best $\delta$ and $\varepsilon$ values are selected for each dataset respectively among the sets $\{n, n/2, n/4, \ldots, n/2^k\}$, with n = max$\{p,q\}$ and $n/2^k \leq .5 < n/2^{k+1}$, and $\{20, 20/2, 20/4, \ldots, 20/2^k\}$, with $20/2^k \leq 1e^{-2} < 20/2^{k+1}$ so as to minimize the classification errors estimated for the **training data**. If different $(\delta, \varepsilon)$ values lead to the minimal error rate estimated for the training data, then the pairs having the **highest** $\delta$ **value are selected first, then the pair with the highest $\varepsilon$ value** is finally selected.

- OTWED: for our experiment, *'stiffness value'* ($\gamma$) is selected from {$1e10^{-5}$, $1e10^{-4}$, $1e10^{-3}$, $1e10^{-2}$, $e10^{-1}$, 1} and $\lambda$ is selected from {0, .25, .5, .75, 1.0 }. The $\gamma$ and $\lambda$ parameter values are selected for each dataset so as to minimize the classification errors estimated on the **training data**. If different ($\gamma, \lambda$) values lead to the minimal error rate estimated for the training data then the pairs containing the **highest $\gamma$ value are selected first, then the pair with the highest $\lambda$ value** is finally selected.

For ERP and OTWED we used the *L1-norm*, while the *L2-norm* has been implemented in DTW and ODTW as reported in [32]. The gap value used in ERP has been set to the distance between the deleted or inserted sample and *0* as suggested by the authors [5].

Finally, as time is not explicitly given for these datasets, we used the index value of the samples as the time stamps for the whole experiment.

This experiment shows that the TWED distance is effective for the considered task comparatively to ED, DTW, ODTW, ERP and LCSS measures, since it exhibits, on average, the lowest error rates for the testing data as shown in *Tab. 1* and *Fig. 3*. The gain, on average, is relatively significant: 2.5% against ODTW, 4.7% against LCSS, 3% against ERP, 9.4% against ED and 8.8% against DTW.

The same experiment carried out for down-sampled time series (*Tab. 2* and *Fig. 4)* shows that the error rates drop more than twice as fast for ED, DTW, ERP, LCSS, ODTW than for TWED. In that experimental context, using time stamps when matching non uniform down-sampled time series seems to be quite effective.

| Dataset | Nbr of classes \| Size of training\|testing set | 1-NN ED | 1-NN ODTW | 1-NN DTW | 1-NN LCSS | 1-NN ERP | 1-NN OTWED |
|---|---|---|---|---|---|---|---|
| **Synthetic Control** | 6\|300\|300 | 0.12 | 0.017 | **0.007** | 0.047 | 0.036 | 0.023 |
| **Gun-Point** | 2\|50\|150 | 0.087 | 0.087 | 0.093 | **0.013** | 0.04 | **0.013** |
| **CBF** | 3\|30\|900 | 0.148 | 0.004 | **0.003** | 0.009 | **0.003** | 0.007 |
| **Face (all)** | 14\|560\|1690 | 0.286 | 0.192 | 0.192 | 0.201 | 0.202 | **0.189** |
| **OSU Leaf** | 6\|200\|242 | 0.483 | 0.384 | 0.409 | **0.202** | 0.397 | 0.248 |
| **Swedish Leaf** | 15\|500\|625 | 0.213 | 0.157 | 0.210 | 0.117 | 0.12 | **0.102** |
| **50Words** | 50\|450\|455 | 0.369 | 0.242 | 0.310 | 0.213 | 0.281 | **0.187** |
| **Trace** | 4\|100\|100 | 0.24 | 0.01 | **0.0** | 0.02 | 0.17 | 0.050 |
| **Two Patterns** | 4\|1000\|4000 | 0.09 | 0.0015 | **0.0** | **0.0** | **0.0** | 0.001 |
| **Wafer** | 2\|1000\|6174 | 0.005 | 0.005 | 0.020 | **0** | 0.009 | 0.004 |
| **Face (four)** | 4\|24\|88 | 0.216 | 0.114 | 0.170 | 0.068 | 0.102 | **0.034** |
| **Lighting2** | 2\|60\|61 | 0.246 | **0.131** | **0.131** | 0.18 | 0.148 | 0.213 |
| **Lighting7** | 7\|70\|73 | 0.425 | 0.288 | 0.274 | 0.452 | 0.301 | **0.247** |
| **ECG** | 2\|100\|100 | 0.12 | 0.12 | 0.23 | **0.100** | 0.13 | **0.100** |
| **Adiac** | 37\|390\|391 | 0.389 | 0.391 | 0.396 | 0.425 | 0.378 | **0.376** |
| **Yoga** | 02\|300\|3000 | 0.170 | 0.155 | 0.164 | 0.137 | 0.147 | **0.130** |
| **Fish** | 7\|175\|175 | 0.267 | 0.233 | 0.267 | 0.091 | 0.12 | **0.051** |
| **Coffee** | 2\|28\|28 | 0.25 | **0.179** | **0.179** | 0.214 | 0.25 | 0.214 |
| **OliveOil** | 4\|30\|30 | 0.133 | 0.167 | **0.133** | 0.8 | 0.167 | 0.167 |
| **Beef** | 5\|30\|30 | 0.467 | **0.467** | 0.5 | 0.533 | 0.5 | 0.533 |
| *MEAN* | | 0.240 | 0.167 | 0.232 | 0.191 | 0.175 | **0.145** |
| *STD* | | 0.131 | **0.136** | 0.145 | 0.210 | 0.140 | 0.139 |

*TAB.1: COMPARATIVE STUDY USING THE UCR DATASETS [16]: CLASSIFICATION ERROR RATE OBTAINED USING THE FIRST NEAR NEIGHBOR CLASSIFICATION RULE FOR ED, DTW, ODTW, LCSS, ERP, AND OTWED DISTANCE*

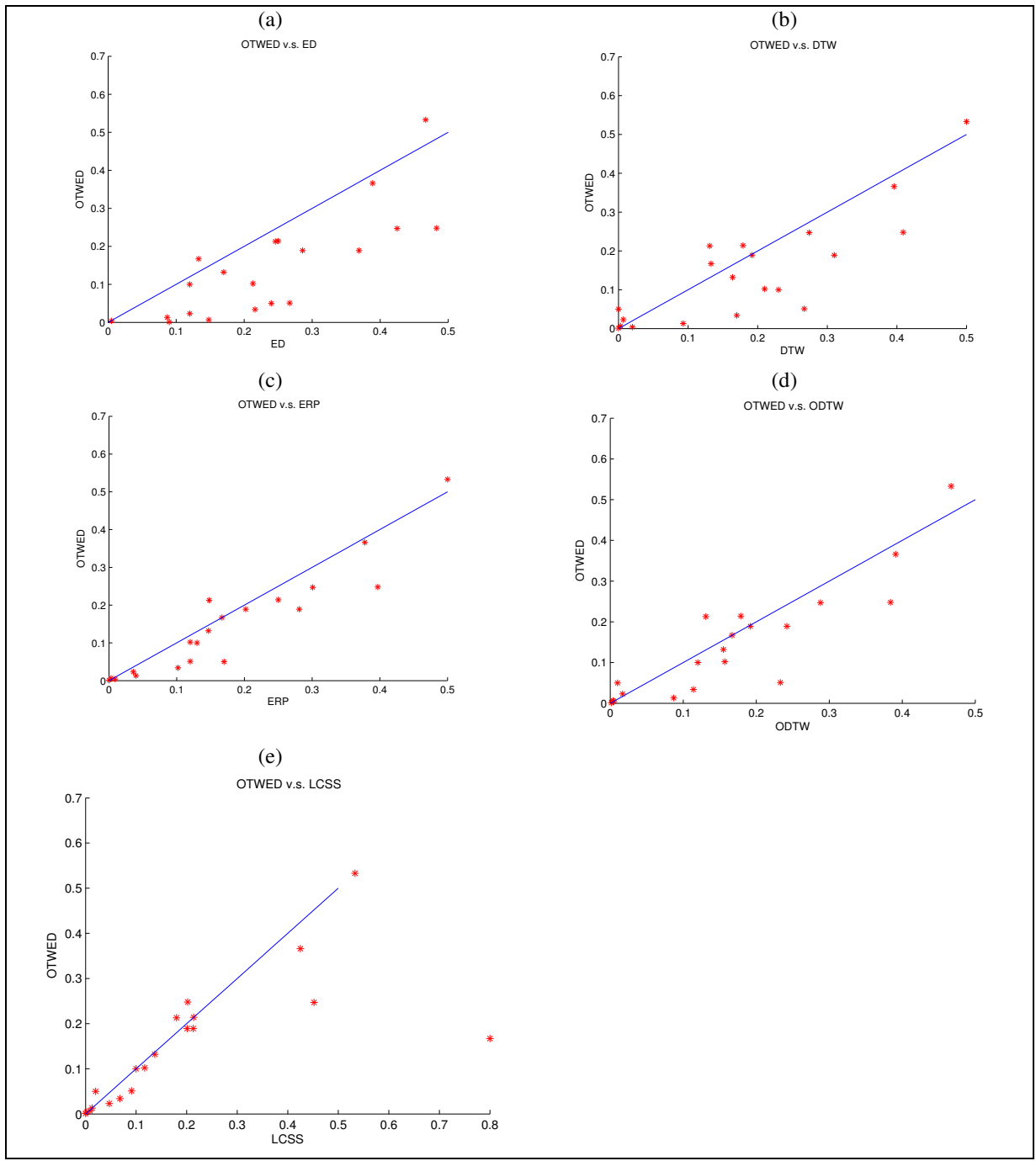

*Fig. 3: Comparison of distance pairs ($\delta x, \delta y$). The x and y axes show the error rates for the two compared distances. The straight line has a slope of 1.0 and dots correspond to the error rate for the selected distance pair and tested data sets. A dot below (resp. above) the straight line indicates that distance $\delta y$ has a lower (resp. higher) error rate than distance $\delta x$. Plot (a) shows OTWED v.s. ED, plot (b) shows OTWED v.s. DTW, plot (c) shows OTWED v.s. ERP, plot (d) shows OTWED v.s. ODTW, plot (e) shows OTWED v.s. LCSS.*

| Dataset | Nbr of classes \| Size of testing set | 1-NN ED | 1-NN ODTW | 1-NN DTW | 1-NN LCSS | 1-NN ERP | 1-NN OTWED |
|---|---|---|---|---|---|---|---|
| **Synthetic Control** | 6\|300\|300 | 0.233 | 0.173 | 0.177 | 0.243 | 0.22 | 0,000 |
| **Gun-Point** | 2\|50\|150 | 0.14 | 0.113 | 0.067 | **0.027** | 0.047 | 0,020 |
| **CBF** | 3\|30\|900 | 0.24 | 0.027 | **0.017** | 0.03 | 0.028 | 0,067 |
| **Face (all)** | 14\|560\|1690 | 0.482 | 0.273 | 0.292 | 0.336 | 0.346 | 0,236 |
| **OSU Leaf** | 6\|200\|242 | 0.541 | 0.455 | 0.43 | 0.393 | 0.475 | 0,281 |
| **Swedish Leaf** | 15\|500\|625 | 0.932 | 0.323 | 0.322 | 0.288 | 0.291 | 0,146 |
| **50Words** | 50\|450\|455 | 0.327 | 0.303 | 0.369 | 0.251 | 0.323 | 0,189 |
| **Trace** | 4\|100\|100 | 0.07 | **0** | **0** | 0 | 0.06 | 0,090 |
| **Two Patterns** | 4\|1000\|4000 | 0.593 | **0** | **0** | 0.104 | 0.013 | 0,001 |
| **Wafer** | 2\|1000\|6174 | 0.025 | 0.014 | 0.022 | 0.018 | 0.013 | 0,010 |
| **Face (four)** | 4\|24\|88 | 0.432 | 0.239 | 0.216 | 0.295 | 0.261 | 0,159 |
| **Lighting2** | 2\|60\|61 | 0.263 | **0.098** | 0.115 | 0.148 | 0.115 | 0,197 |
| **Lighting7** | 7\|70\|73 | 0.521 | 0.315 | 0.342 | 0.427 | **0.26** | 0,370 |
| **ECG** | 2\|100\|100 | 0.18 | 0.25 | 0.26 | 0.32 | 0.2 | 0,110 |
| **Adiac** | 37\|390\|391 | 0.527 | 0.486 | 0.483 | 0.448 | 0.496 | 0,417 |
| **Yoga** | 02\|300\|3000 | 0.204 | 0.166 | 0.171 | 0.188 | 0.189 | 0,140 |
| **Fish** | 7\|175\|175 | 0.371 | 0.354 | 0.354 | **0.189** | 0.28 | 0,086 |
| **Coffee** | 2\|28\|28 | 0.179 | **0.143** | 0.179 | 0.214 | 0.25 | 0,285 |
| **OliveOil** | 4\|30\|30 | 0.567 | 0.167 | 0.167 | 0.333 | 0.333 | 0,167 |
| **Beef** | 5\|30\|30 | 0.533 | 0.533 | 0.5 | 0.5 | 0.5 | 0,333 |
| *MEAN* | | 0,37 | 0,22 | 0,22 | 0,24 | 0,24 | 0,165 |
| *STD* | | 0,22 | 0,16 | 0,16 | 0,15 | 0,16 | 0,124 |

*TAB.2: COMPARATIVE STUDY USING THE UCR DATASETS [16]: CLASSIFICATION ERROR RATE OBTAINED USING THE FIRST NEAR NEIGHBOR CLASSIFICATION RULE ON <u>DOWN-SAMPLED</u> TIME SERIES FOR ED, DTW, ODTW, LCSS, ERP, AND OTWED DISTANCE*

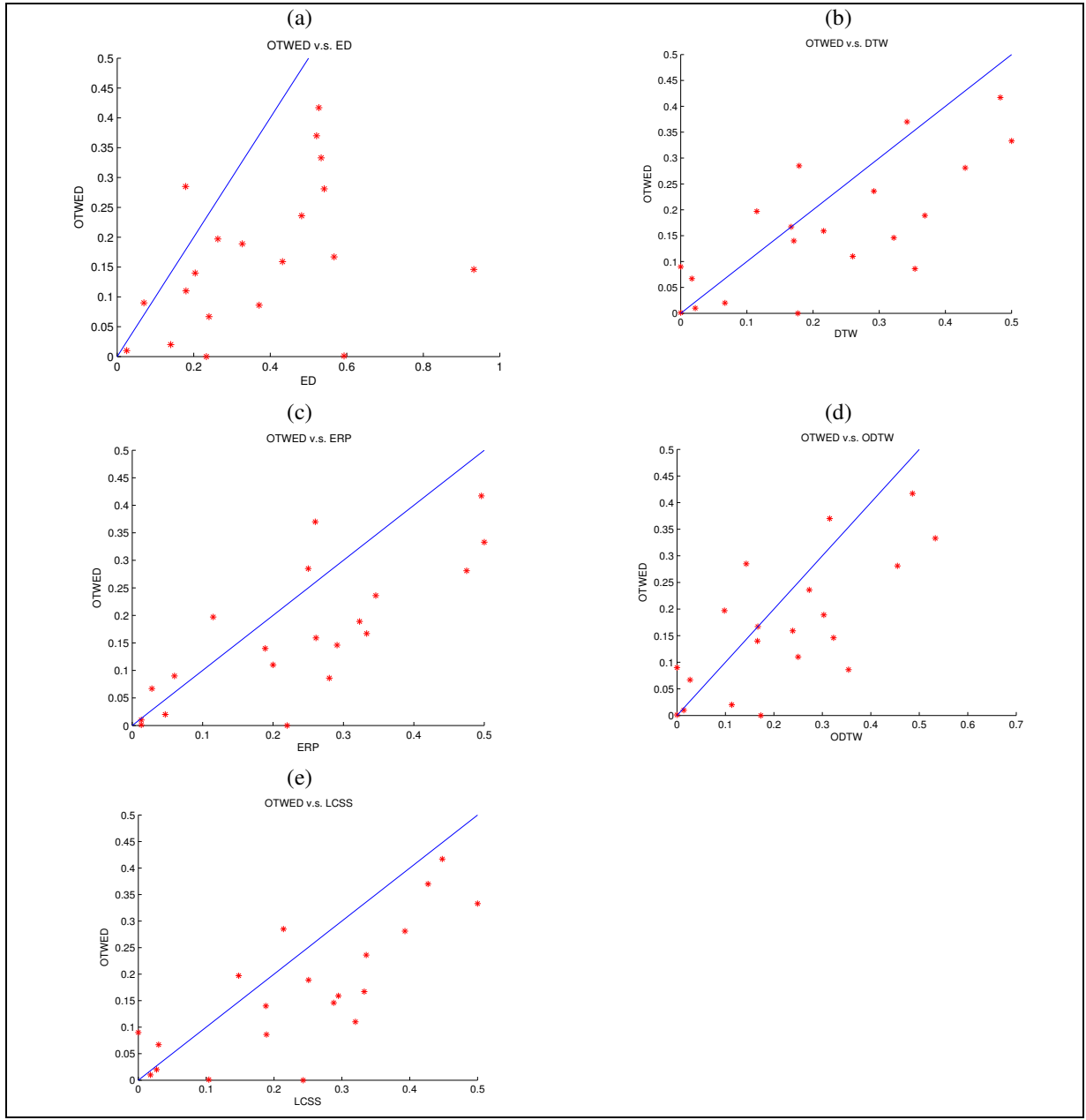

*Fig. 4: Comparison of distance pairs (δx, δy) <u>applied on down sampled times series</u>. The x and y axes show the error rates for the two compared distances. The straight line has a slope of 1.0 and dots correspond to the error rate for the selected distance pair and tested data sets. A dot below (resp. above) the straight line indicates that distance δy has a lower (resp. higher) error rate than distance δx. Plot (a) shows OTWED v.s. ED, plot (b) shows OTWED v.s. DTW, plot (c) shows OTWED v.s. ERP,  plot (d) shows OTWED v.s. ODTW, plot (e) shows OTWED v.s. LCSS.*

B. *Range query search experiment*

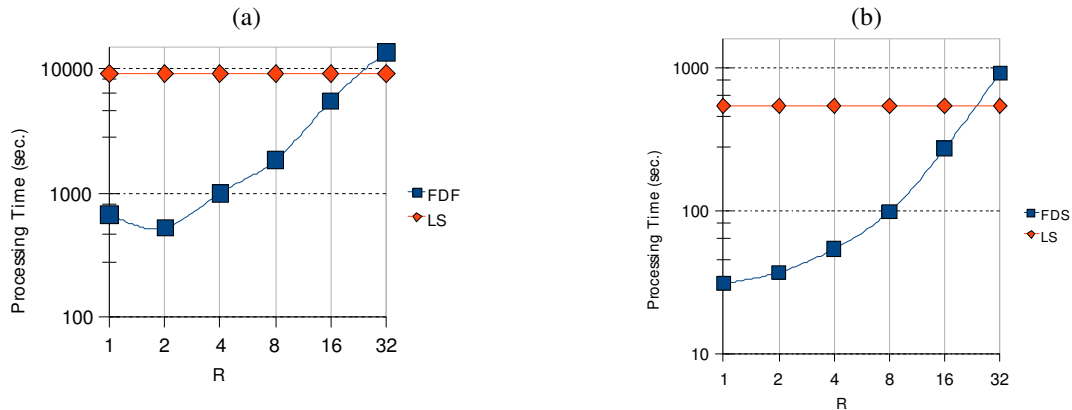

*Fig. 5: Processing time (in seconds) required to filter 100 random queries as a function of the radius R for (a) the heterogeneous database and (b) the homogeneous database. The rhombuses constant line refers to the Linear Scanning (LS) procedure, while the square line refers to the FDS procedure.*

Using the multiresolution approach defined in [24] to get nested PCWA approximations of time series in linear time complexity, the Fast and Dirty Filter (FDF) that we propose iteratively evaluates the inequality (13) from the crudest level of resolution to the finest (which corresponds to the original time series). Between two successive levels of resolution, halves of the samples are eliminated. Given a radius $R$ and a reference time series $A$, the FDF rejects candidate time series $B$ as early as possible, i.e. as soon as two approximations of $A$ and $B$ satisfy the inequality (13). The experiment consists in evaluating the processing time required to extract all the times series $B$ located inside a ball of radius $R$ centered on the reference time series $A$ which is drawn randomly from a database. We compare the FDF against a Linear Scanning (LS) procedure applied in the original time series space.

The first experiment is carried out from a heterogeneous database composed of the 20 datasets available at UCR [16]. This database comprises *23999* time series. The second experiment is carried out from a homogeneous database which is composed only of the *Two_patterns* dataset available at UCR [16]. This database comprises *5000* time series. For both experiments $\lambda$ and $\gamma$ parameters are set constant equal to the intermediate value $0.01$.

Figure 5 shows that the FDF performs well for small radius for both databases. The FDF is by an order of magnitude faster than the LS procedure for radius varying from *1* to *4*. The FDF matches LS for a radius *R* in between values *16* and *32*. The FDF is performing worse than LS for greater radius, mainly because the inequality (13) does not efficiently apply anymore.

## VII. CONCLUSION

From a graphical curve editing perspective and from earlier work on symbolic edit distance and dynamic time warping we have developed an elastic similarity measure called TWED to match time series with some time shifting tolerance. We have proved that the TWED measure is a metric, and as such TWED can be used complementarily with methods developed for searching in metric spaces as potential solutions for time series searching and retrieval applications when time shift tolerance is concerned. The originality of TWED, comparatively to similar elastic measures, apart from the way insertions and deletions are managed, lies in the introduction of a parameter which controls the '*stiffness*' of the measure thus placing TWED in between the Euclidian distances (infinite stiffness) and the DTW similarity measure (null stiffness). Moreover TWED involved a second parameter which defines a constant penalty for insert or delete operations, similarly to the edit distance defined for string matching. These two parameters can be straightforwardly optimized for each application or dataset as far as training data are available.

Furthermore, a procedure has been drawn up to lower bound the TWED metric. This procedure consists in approximating the time series using polygonal or piecewise constant approximations. It takes benefits from the triangle inequality to link the TWED measure evaluated on the approximated representations of time series to the TWED measure evaluated on the original time series. The computational cost reduction of TWED when evaluated in the approximated representation space is quadratic with the compression rate of the approximation. Nevertheless, this kind of lower bound has no linear complexity. Experimentation shows that one can expect to gain an order of magnitude in processing time using a fast and dirty filter based on this lower bound. The search for a lower bound whose complexity is effectively linear and that could be efficiently used in conjunction with down-sampled approximation of time series is still a perspective.

The empirical quality of the distance has been evaluated through a classification experiment based on the first near neighbor classification rule for *20* different datasets. Globally, for this experiment, TWED performs, on average, significantly better than the Euclidian distance and Dynamic Time Warp measure and slightly better than the Longest Common Subsequence measure, the Edit Distance with Real Penalty and the Dynamic Time Warping measure with optimized search corridor size. When the classification experiment is applied to down-sampled time series, TWED is more robust than the other tested measures. This is mainly because the times series are not uniformly sampled in this experiment in which case it is relevant for time stamps.

## VIII. APPENDIX

From propositions 1 to 4 we can upper-bound the matching of two time series evaluated in the original space with the matching of their approximations evaluated in the down-sampled space as follows:

$\forall \lambda \geq 0, \gamma > 0, \quad \forall A_1^p, B_1^q \in U^2, \ \forall r \in [1; p[, \ \forall r' \in [1; q[ \text{ we have}:$

$\delta_{\lambda,\gamma}(\overline{A}_1^{p,r}, A_1^p) \leq 2 \cdot D_{LP}(\overline{A}_1^{p,r}, A_1^p);$ since $\overline{A}_1^{p,r}$ and $A_1^p$ have the same length

$\delta_{\lambda,\gamma}(\overline{A}_1^{p,r}, \tilde{A}_1^r) \leq \lambda \cdot (p - r) + \gamma \cdot \Delta T(2 \cdot p - r);$ Proposition 4

$\delta_{\lambda,\gamma}(\overline{B}_1^{q,r'}, B_1^q) \leq 2 \cdot D_{LP}(\overline{B}_1^{q,r'}, B_1^q);$ since $\overline{B}_1^{q,r'}$ and $B_1^q$ have the same length

$\delta_{\lambda,\gamma}(\overline{B}_1^{q,r'}, \tilde{B}_1^{r'}) \leq \lambda \cdot (q - r') + \gamma \cdot \Delta T(2 \cdot q - r');$ Proposition 4

From these inequalities and the triangle inequality verified by $\delta_{\lambda,\gamma}$ we get:

$$\delta_{\lambda,\gamma}(\tilde{A}_1^r, A_1^p) \leq \lambda \cdot (p - r) + \gamma \cdot \Delta T(2 \cdot p - r) + 2 \cdot D_{LP}(\overline{A}_1^{p,r}, A_1^p)$$

$$\delta_{\lambda,\gamma}(\tilde{B}_1^{r'}, B_1^q) \leq \lambda \cdot (q - r') + \gamma \cdot \Delta T(2 \cdot q - r') + 2 \cdot D_{LP}(\overline{B}_1^{q,r'}, B_1^q)$$

And since $\delta_{\lambda,\gamma}$ verifies the triangle inequality we have:

$\delta_{\lambda,\gamma}(\tilde{A}_1^r, \tilde{B}_1^{r'}) \leq \delta_{\lambda,\gamma}(\tilde{A}_1^r, A_1^p) + \delta_{\lambda,\gamma}(A_1^p, B_1^q) + \delta_{\lambda,\gamma}(B_1^q, \tilde{B}_1^{r'})$
$\quad \leq \lambda \cdot (p + q - r - r') + \gamma \cdot \Delta T(2 \cdot (p + q) - r - r') +$
$\quad \quad 2 \cdot D_{LP}(\overline{A}_1^{p,r}, A_1^p) + 2 \cdot D_{LP}(\overline{B}_1^{q,r'}, B_1^q) + \delta_{\lambda,\gamma}(A_1^p, B_1^q)$

$\delta_{\lambda,\gamma}(A_1^p, B_1^q) \leq \delta_{\lambda,\gamma}(A_1^p, \tilde{A}_1^r) + \delta_{\lambda,\gamma}(\tilde{A}_1^r, \tilde{B}_1^{r'}) + \delta_{\lambda,\gamma}(\tilde{B}_1^{r'}, B_1^q)$
$\quad \leq \lambda \cdot (p + q - r - r') + \gamma \cdot \Delta T(2 \cdot (p + q) - r - r') +$
$\quad \quad 2 \cdot D_{LP}(\overline{A}_1^{p,r}, A_1^p) + 2 \cdot D_{LP}(\overline{B}_1^{q,r'}, B_1^q) + \delta_{\lambda,\gamma}(A_1^p, B_1^q)$

Leading to:

$$\left| \delta_{\lambda,\gamma}(A_1^p, B_1^q) - \delta_{\lambda,\gamma}(\tilde{A}_1^r, \tilde{B}_1^{r'}) \right| \leq \lambda \cdot (p + q - r - r') + \gamma \cdot \Delta T(2 \cdot (p + q) - r - r') +$$
$$2 \cdot D_{LP}(\overline{A}_1^{p,r}, A_1^p) + 2 \cdot D_{LP}(\overline{B}_1^{q,r'}, B_1^q)$$